\documentclass[oneside, 12pt]{amsart} 
\usepackage{amssymb, amsmath, latexsym, fancyhdr}
\usepackage{epsfig}
\usepackage{graphicx}

\newtheorem{theorem}{Theorem}[section]
\theoremstyle{definition}

\theoremstyle{remark}
\newtheorem{remark}[theorem]{Remark}

\newcommand{\bi}{\begin{itemize}}
\newcommand{\ei}{\end{itemize}}

\newcommand{\ZZ}{\mathbb{Z}}

\newcommand{\datum}{March 25, '13}

\date{\datum}

\address{Corresponding author: Steven P. Ellis \\
Unit 42, New York State Psychiatric Institute at Columbia University \\
1051 Riverside Dr. \\
New York, NY 10032 \\
U.S.A.\\
e-mail: spe4@columbia.edu}

\keywords{dichotomous data, high-order dependence, Fourier analysis of time series, computational homology, persistent homology, fMRI, ADHD}

\thanks{\emph{2010 AMS Subject Classification.}  Primary: 62H17; Secondary: 92C55, 62M15}

\thanks{This research is supported in part by United States PHS grants MH62185 and MH084029}

\begin{document}

\title[Concurrence Topology and {fMRI}]{Describing High-Order Statistical Dependence Using ``Concurrence Topology'', with Application to Functional MRI Brain Data}
       
\author{Steven P.\ Ellis and Arno Klein}

\maketitle

\begin{abstract}
For multivariate data, dependence beyond pair-wise can be important. This is true, for example, in  using functional MRI (fMRI) data to investigate brain functional connectivity. \linebreak
When one has more than a few variables, however, the number of simple summaries of even third-order dependence can be unmanageably large.

``Concurrence topology'' is an apparently new nonparametric method for describing high-order dependence among up to dozens of dichotomous variables (e.g., seventh-order dependence in 32 variables). This method generally produces summaries of $p^{th}$-order dependence of manageable size no matter how big $p$ is. (But computing time can be lengthy.) For time series, this method can be applied in both the time and Fourier domains.

Write each observation as a vector of 0's and 1's. A ``concurrence'' is a group of variables all ``1'' in the same observation. The collection of concurrences can be represented as a sequence of shapes (``filtration'').

Holes in the filtration indicate weak or negative association among the variables. The pattern of the holes in the filtration can be analyzed using computational topology. 

This method is demonstrated on dichotomized fMRI data. The dataset includes subjects diagnosed with ADHD and healthy controls. In an exploratory analysis numerous group differences in the topology of the filtrations are found.
\end{abstract}

\section{Introduction} \label{S:intro}
We propose an apparently new nonparametric method, ``concurrence topology'', for describing the high-order dependence structure of multivariate binary data. It does this by translating the data into a series of shapes and then analyzing the topology of those shapes. 
In this paper our main focus is a specific version concurrence topology we call ``concurrence homology''.

Concurrence topology can be applied to the population joint distribution. But this paper is about concurrence topology as a descriptive method applied to data. However, standard inferential statistical methods can be applied to perform inferences based on subject-wise descriptions produced by concurrence topology. That is the approach we take here (section \ref{S:data.analysis}). 

Concurrence topology was initially developed for analysis of ``functional connectivity'' in resting-state functional magnetic resonance \linebreak
imaging data (fMRI, \cite{pJpmMsmS02.fMRIbook},  \linebreak
\cite{VandenHeuvel2010}). We use such data as a test bed for the method.    
This data set consists of multivariate time series of ``blood oxygen level dependent (BOLD)'' values for each of 25 patients diagnosed with attention deficit hyperactivity disorder (ADHD), 
and 41 healthy controls (section \ref{S:fMRI.data}). (Concurrence topology applies to binary data so we first dichotomized the fMRI BOLD time series values, section \ref{S:dichotomization}.) Others have used fMRI to reveal abnormalities in functional connectivity in ADHD
\cite{yPmaMjKaP07.fMRI_ADHD}. We find other differences using concurrence topology.
 
A binary variable $X$ can be thought of as taking values in the set $\{ 0, 1 \}$. With a nod to fMRI terminology, say that $X$ is ``active'' when it is ``1''.  
Informally, variables $X_{1}, \ldots, X_{p}$ are ``positively associated'' if, when some of the variables are active, all $p$ variables tend to be active. 

Concurrence homology is sensitive to weak or negative, i.e., nonpositive, association.  
Thus, variables $X_{1}, \ldots, X_{p}$ are weakly or negatively associated if, compared to the number of times (frequency) at which \emph{some} of them are active, the frequency at which they are \emph{all} active is low. The frequency at which some of them are active can be checked by looking at fewer than $p$ variables at a time. Similarly, the definition of  the interaction term $\lambda^{ X_{1}, \ldots, X_{p} }_{11\ldots1}$ in a log linear model 
\cite[p.\ 143]{aA90.CatDatAnal} involves not just the product $X_{1} \ldots X_{p}$ but also lower-dimensional marginals.
 
If a feature of the joint distribution can be detected by looking at $p$ variables at a time, but not by looking only at $p-1$ variables at a time, then we say that feature has to do with the ``$p^{th}$-order dependence'' among the variables. For example, Pearson, Kendall, and Spearman correlation are measures of $2^{nd}$-order dependence because a correlation matrix for a collection of variables can be computed by looking at the variables two at a time. The odds ratio is also second-order. In this paper we focus on ``high-order'' dependence, by which we mean dependence of order at least three. 

Assuming \emph{a priori} structure for the dependence among the variables or designating \emph{a priori} some variables as ``predictors'' and others as ``responses'' can be a powerful way to learn from data. However, our interest is in ``agnostic'' methods. A method is ``agnostic'' if \emph{a priori}, for $k = 1, 2, \ldots$ all groups of $k$ variables are treated identically. This rules out  much \emph{a priori} structural assumptions. An example of an agnostic method is principal component analysis, a second-order method.

There are apparently few nonparametric agnostic methods that can cope with the ``combinatorial explosion'' (section \ref{S:combo.explosion}) that is inherent in describing high-order dependence among more than a few variables. Other such methods include independent component analysis (ICA, 
\cite{aHjKeO01.IndepCompAnlys}), latent variable methods \linebreak
\cite{dBmKiM11.LatentVariables}, and, perhaps, the method of \linebreak
\cite{dDcX09.BayesCategoricalDependence}. (There is a large literature on analysis of fMRI data, e.g., 
\cite{fgA11.StatAnlysfMRI,Li2009}, \linebreak
and \cite{VandenHeuvel2010}.)

The aforementioned methods and ours capture very different aspects of high-order dependence. Hence, \emph{prima facie} these methods are not competitors. For that reason, and in the interest of brevity, in this paper we do not compare concurrence topology to other methods.

\section{Toy examples}  \label{S:toy.story}
Concurrence homology is based on ideas from algebraic topology 
\cite{jrM84,hEjlH10.CompTopol},
but can be ex- \linebreak
plained nontechnically via ``toy examples''. Consider the three multivariate datasets shown in table \ref{Ta:same.to.2nd.not.to.3rd}. Each has five variables. The univariate marginal distributions are the same across all three datasets. The bivariate marginals are also the same. But the datasets differ in third-order. 
For example, in dataset I $X$, $Y$, and $Z$ are never all active in the same row, but in the other two datasets they sometimes are. 

Concurrence topology is based on ``concurrences''. A concurrence is a group of variables that are all active in the same observation. In effect, we throw away the 0's and just retain the 1's. (So if an observation consists entirely of 0's, it is dropped.) Call the number of variables in the concurrence the ``length'' of the concurrence.

   \begin{table}[h]
	\begin{tabular}{| p{.15in} | p{.15in} | p{.15in} | p{.15in} | p{.15in} || p{.15in} | p{.15in} | p{.15in} | 
	             p{.15in} | p{.15in} || p{.15in} | p{.15in} p{.15in} | p{.15in} | p{.15in} | p{.15in}  }
	   \multicolumn{5}{c||}{I} & \multicolumn{5}{c||}{II} & \multicolumn{5}{c}{III} \\
$V$ & $W$ & $X$ & $Y$ & $Z$ & $V$ & $W$ & $X$ & $Y$ & $Z$ & $V$ & $W$ & $X$ & $Y$ & $Z$ 
                        \\ \hline
0 & 0 & 0 & 0 & 0 & 0 & 0 & 0 & 0 & 0 & 0 & 0 & 0 & 0 & 1 \\
0 & 0 & 0 & 1 & 1 & 0 & 0 & 0 & 0 & 1 & 0 & 0 & 0 & 1 & 0 \\
0 & 0 & 1 & 0 & 1 & 0 & 0 & 0 & 1 & 0 & 0 & 0 & 1 & 0 & 0 \\
0 & 0 & 1 & 1 & 0 & 0 & 0 & 1 & 0 & 0 & 0 & 0 & 1 & 1 & 1 \\
0 & 0 & 0 & 0 & 0 & 0 & 0 & 0 & 1 & 1 & 0 & 0 & 0 & 0 & 1 \\
0 & 0 & 0 & 1 & 1 & 0 & 0 & 1 & 0 & 1 & 0 & 0 & 0 & 1 & 0 \\
0 & 0 & 1 & 0 & 1 & 0 & 0 & 1 & 1 & 0 & 0 & 0 & 1 & 0 & 0 \\
0 & 0 & 1 & 1 & 0 & 0 & 0 & 1 & 1 & 1 & 0 & 0 & 1 & 1 & 1 \\
1 & 0 & 0 & 0 & 1 & 1 & 0 & 0 & 0 & 1 & 1 & 0 & 0 & 0 & 1 \\
0 & 1 & 1 & 0 & 0 & 0 & 1 & 1 & 0 & 0 & 0 & 1 & 1 & 0 & 0 \\
1 & 1 & 0 & 0 & 0 & 1 & 1 & 0 & 0 & 0 & 1 & 1 & 0 & 0 & 0 \\
	\end{tabular}
\caption{Three datasets identical up to second-order, but not at third-order.}
\label{Ta:same.to.2nd.not.to.3rd}
\end{table}

For example, the ``concurrence list'' in data set I in table \ref{Ta:same.to.2nd.not.to.3rd} is $YZ$, $XZ$, $XY$, $YZ$, $XZ$, $XY$, $VZ$, $WX$, and $VW$. (The first row is dropped. We ignore the order, but not the frequency of appearance, of concurrences.)

We represent a concurrence list as a shape. In general, the shape will not fit on a plane or in three-dimensional space. But the shapes corresponding to data sets I, II, and III do fit on a plane. Choose points (``vertices'') on the plane, each corresponding to a variable. Be careful that no three points fall fall on the same line. If two variables form a concurrence in the list then connect the corresponding vertices by a line segment (``1-simplex''). If three variables are concurrent, connect them by a solid triangle (``2-simplex''). (Of course, if three variables are concurrent then each pair of them are. Implicit in connecting three variables by a 2-simplex is connecting each pair by a 1-simplex.) Vertices may need to be rearranged so that none fall in the middle of a 2-simplex. 

\begin{figure}
      \epsfig{file = 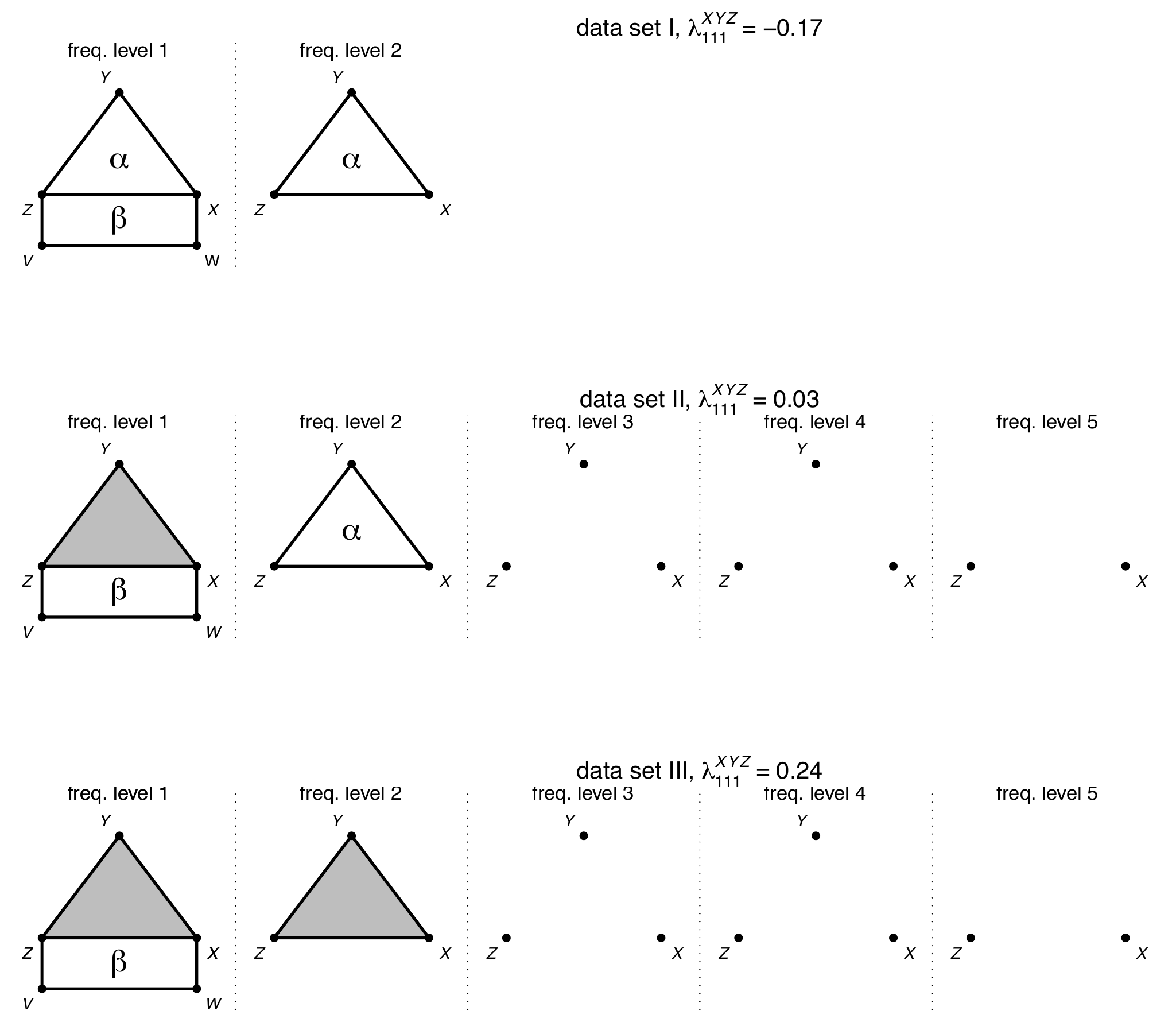, height = 3.8in, , }  
      \caption{Rows are filtered Curto-Itskov complexes for data sets in table \ref{Ta:same.to.2nd.not.to.3rd}. Columns, separated by dotted vertical lines, correspond to frequency levels. ``$\lambda_{111}^{XYZ}$'' is a third-order interaction in a log linear model. ``$\alpha$'' and ``$\beta$'' label holes.}  
      \label{F:toy.filtered.conc.cmplxes}
\end{figure}

We call this shape the ``Curto-Itskov complex'' of the concurrence list. (Strictly speaking, a Curto-Itskov complex is more than just a shape. It is a ``simplicial complex'', a collection of simplices. The name ``Curto-Itskov'' is explained presently.) The Curto-Itskov complexes of the data in table \ref{Ta:same.to.2nd.not.to.3rd} are shown in the first column of figure \ref{F:toy.filtered.conc.cmplxes}. 

The shapes in the first column of figure \ref{F:toy.filtered.conc.cmplxes} distinguish data set I from data sets II and III, but do not distinguish data sets II and III from each other. To do this we construct a decreasing series of shapes, indexed by ``frequency level'', which is how often a concurrence appears in a concurrence list, $\mathcal{C}$.

We say that a concurrence in $\mathcal{C}$ appears even if it appears as part of a larger concurrence.  Thus, in data set II the concurrence $XYZ$ appears once, $XY$ appears twice, and $X$ appears five times. The shape (Curto-Itskov complex), or ``frame'', in frequency level $f$ is constructed in the manner we just described but from the concurrence list, $\mathcal{C}_{f}$ that consists only of concurrences in $\mathcal{C}$ that \emph{appear at least $f$ times.} (In the fMRI data, each time series has the same length, 192. This allows us to use absolute, i.e., integer frequencies, $f$. In general, relative, i.e., fractional, frequencies must be used. A population version of the filtered Curto-Itskov complex might be indexed by a continuum of frequencies.) 
We call this series of shapes the ``filtered Curto-Itskov complex'' of the data. (Conventionally in topology a filtered simplicial complex would be indexed in the opposite order. But in concurrence topology \emph{descending} filtrations are more natural.)

In figure \ref{F:toy.filtered.conc.cmplxes} each row is the filtered Curto-Itskov complex for a data set in table \ref{Ta:same.to.2nd.not.to.3rd}. We see that the filtered Curto-Itskov complexes do distinguish the three data sets. (In fact, the filtered Curto-Itskov complex for a data set together with the number of observations is equivalent to the contingency table for the data set.)

Our work on concurrence topology is inspired by Curto and Itskov \linebreak
\cite{cCvI08.CellGroupsHomol}, who investigated a question in theoretical neuroscience by applying topological methods to simulated data. From each simulation Curto and Itskov constructed a single shape, like those in the first column of figure \ref{F:toy.filtered.conc.cmplxes}, and studied the holes in that shape. For their purpose it was not necessary to build a filtration, i.e., a series of shapes. In essence, they only needed to know whether each cell in a contingency table was 0 or not. But typically for data analysis one needs to know the actual values in the table. To represent those values geometrically a single Curto-Itskov complex is not sufficient.

We call investigation of the joint distribution of multivariate dichotomous data by analyzing the topology of the corresponding filtered Curto-Itskov complex ``concurrence topology''.

\section{Holes} \label{S:holes}
One topological feature of a filtered Curto-Itskov complex is the pattern of holes in the frames. Concurrence homology is a form of concurrence topology that describes that pattern. The main principle of our approach is that holes in filtered Curto-Itskov complexes represent negative or weak association among the variables. (Thus, concurrence homology does not provide a complete description of dependence.) The representation takes into account lower-order dependence. 

One kind of hole is a gap. Viewing cluster analysis as a method for finding, not clusters, but gaps between the clusters, then concurrence homology can be thought of as ``single linkage cluster analysis on steroids'' 
\cite{bsEsLmLdS11.ClusterAnalysis}.
 
Figure \ref{F:toy.filtered.conc.cmplxes} also displays the values of the third-order interaction term $\lambda_{111}^{XYZ}$ in a log linear model for each data set. This term pertains to the frequency of the event $X=1, Y=1, Z=1$. (The contingency tables for these data sets contain 0-cells. For that reason we added 1/2 to each cell in the tables before computing $\lambda_{111}^{XYZ}$, 
\cite[p.\ 137]{aA90.CatDatAnal}.)

Notice that there is a perfect negative association between the number of empty triangles with vertices labeled $X$, $Y$, and $Z$ (labeled ``$\alpha$'') and the values of $\lambda_{111}^{XYZ}$. Thus, it seems that the number of these empty triangles does indicate how negative or weak is the third-order dependence among $X$, $Y$, and $Z$. We do not claim that there will always be such a neat pattern, but it does provide ``experimental evidence'' in favor of our contention that holes in a filtered Curto-Itskov complex indicate weak or negative association. We provide a general argument for that contention in section \ref{S:more.cases.vars}.
 
Now consider the rectangular holes ``$\beta$'' in figure \ref{F:toy.filtered.conc.cmplxes} with vertices $V$, $W$, $X$, and $Z$. Just as an empty triangle pertains to third-order dependence an empty rectangle pertains to fourth-order dependence. However, a fuller picture comes from considering the \emph{dimension} of holes.

A loop of wire represents a 1-dimensional hole because a length of wire is 1-dimensional. The void inside a basketball is 2-dimensional because the basketball consists of sheets of rubber glued together and sheets are 2-dimensional shapes. (Determining the dimension of a hole is not always as straightforward as these examples suggest.) 

\begin{figure}
      \epsfig{file = 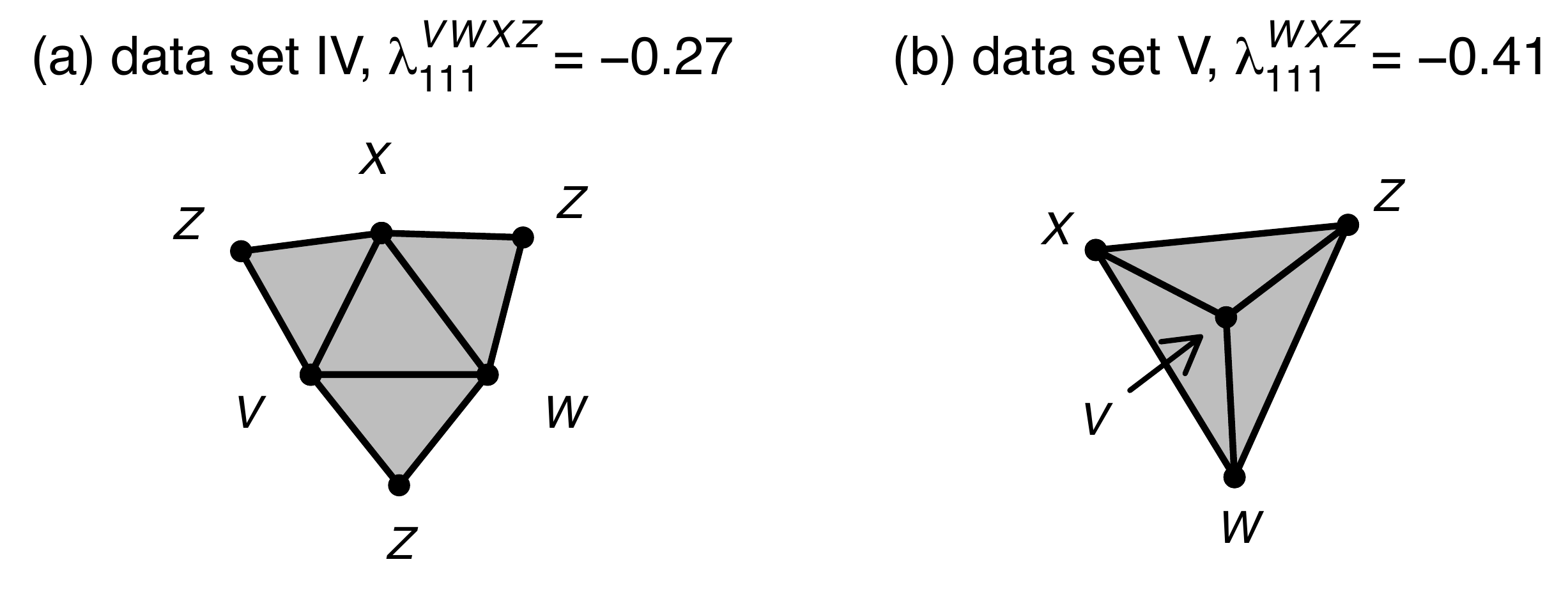, width = 4.5in, }  
      \caption{Curto-Itskov complexes for data sets IV and V.}  
      \label{F:DataSetIV_mock_up}
\end{figure}

To illustrate, consider data set IV listed in table \ref{Ta:hollow.tetrahedron.data}.
   \begin{table}[h]
	\begin{tabular}{| p{.15in} | p{.15in} | p{.15in} | p{.15in} | }
$V$ & $W$ & $X$ & $Z$ \\ \hline
  0 & 1 & 1 & 1 \\
  1 & 0 & 1 & 1 \\
  1 & 1 & 0 & 1 \\
  1 & 1 & 1 & 0 \\
	\end{tabular}
\caption{Data set IV: A data set with weak fourth-order dependence.  } 
\label{Ta:hollow.tetrahedron.data}
\end{table} 
Figure \ref{F:DataSetIV_mock_up}(a) shows the filtered Curto-Itskov complex for data set IV. In this case the filtered Curto-Itskov complex includes just one frame, which forms a hollow tetrahedron, a pyramid with a triangular base. (To represent it as a plane figure it has been opened up. To restore the Curto-Itskov complex, imagine folding the figure along each of the lines $VW$, $VX$, and $WX$ to bring the three $Z$'s into coincidence.) 

The sides of the shape are two-dimensional and enclose a two-\linebreak
dimensional hole. For this data set the fourth-order term $\lambda_{1111}^{VWXZ}$ is -0.27, while the third-order terms $\lambda_{111}^{VWX}$, etc., are all -0.14 so, empirically, the shape seems to encode $4^{th}$-order dependence.

A hole in a Curto-Itskov complex is a global feature of the joint distribution. I.e., it involves \emph{all} the variables. For example, consider data set V, obtained by dropping the first observation (row) in data set IV. Its filtered Curto-Itskov complex contains only one frame (figure \ref{F:DataSetIV_mock_up}(b)). 
For the moment ignore the variable $V$. Then data set V includes the concurrences $XZ$, $WZ$, and  $WX$, but not $WXZ$. 
Thus, the simplices involving only $W$, $X$, and $Z$ form an empty triangle. That, however, is not a hole in the complex because the 2-simplices $VWX$, $VWZ$, and $VXZ$ fill in the empty triangle and concurrence homology does not ignore variables. The fact that holes involve all the variables means that there are usually few holes in filtered Curto-Itskov complexes. This furthers the goal of parsimony (section \ref{S:combo.explosion}). Now, for data set V, $\lambda^{WXZ}_{111} = -0.41$, which is fairly far from 0. Nonetheless, concurrence homology finds no evidence of weak or negative dependence in this data set.

\section{More cases and more variables}  \label{S:more.cases.vars}
In sections \ref{S:fMRI.data}, \ref{S:data.analysis}, \ref{S:some.findings}, and \ref{SS:localization.in.fMRI.data} we consider real brain imaging data sets that have 32 or 74 variables (after some variables have been dropped, section \ref{S:dichotomization}). In the latter case it is common for 60 or more variables to be active in a single time point for a single subject. (We analyzed the concurrence topology for each subject separately. So for concurrence topology purposes we had 66 data sets, not one. We then studied the distributions of summaries thereof across subjects, section \ref{S:data.analysis}.)

We can still apply our approach to such data sets. This can be done abstractly, but as an aid to intuition we start by imagining that the variables correspond to points (``vertices'') in general position (for any $k = 1, 2, \ldots$ no set of $k$ points lie on a plane of dimension $k-2$) in a high-dimensional space.  

There are simplices of any dimension 
\cite[\S 1]{jrM84}. We already observed that a line segment and a solid triangle are simplices of dimensions 1 and 2, respectively. A 3-simplex is a \emph{solid} tetrahedron. (A 0-simplex is a single point.) A simplex is completely determined by its vertices (``corners''). 

For any collection of concurrent variables, even 60 or more, insert into the high-dimensional space the simplex whose vertices correspond to the concurrent variables. We call the resulting collection of simplices the ``Curto-Itskov complex'' of the data set. Considering multiplicity of concurrences as in section \ref{S:toy.story}, one gets a series of shapes, the ``filtered Curto-Itskov complex''.

One cannot detect a $d$-dimensional hole, $\eta$, in a Curto-Itskov complex by only looking at $d+1$ variables at a time, but one can detect a $d$-dimensional hole by looking at groups of $d+2$ variables at a time. Detection of the hole may require looking at multiple groups of $d+2$ variables at the same time (e.g., the holes ``$\beta$'' in figure \ref{F:toy.filtered.conc.cmplxes}). Thus, $\eta$ reflects dependence of order $d+2$ or higher. 

Moreover, the hole $\eta$ is bounded by at least $d+2$ simplices, each corresponding to $d+1$ variables active at the same time. However, for $\eta$ to exist also requires one or more groups of $d+2$ of the same variables to \emph{not} be active at the same time. Thus, existence of $\eta$ reflects a shortage of active groups of $d+2$ variables compared to active groups of $d+1$ variables. To sum up: 
	\begin{multline} \label{E:dim.d.order.d+2}
		\text{A $d$-dimensional hole in a filtered Curto-Itskov} \\ 
		        \text{complex indicates relatively weak or negative} \\
		                \text{association of order } d+2 \text{ or higher.}
	\end{multline}
 
 
\begin{remark} \label{R:high.dim.holes}
For high $d$, a $d$-dimensional hole in a filtered Curto-Itskov complex actually indicates strong \emph{absolute} association because a $d$-dimensional hole is bounded by at least $d+2$, $d$-simplices, each corresponding to one or more concurrences of length $d+1$. Strong, but not perfect, association is needed to generate so many long concurrences. This is exemplified in section \ref{SSS:dim2.WB.F.domn}.
\end{remark}
  

``Homology theory'' is a branch of algebraic topology that is concerned with tunnels, holes, voids, cavities, etc., in shapes. \linebreak 
\cite{jrM84,hEjlH10.CompTopol}.
The homologically \linebreak
correct term for ``hole'' is ``homology class''. 
In this paper we use homology (with $\ZZ/2 = \{ 0, 1 \}$ coefficients) to describe the patterns of holes in filtered Curto-Itskov complexes.  We call this approach to concurrence topology ``concurrence homology''. 

We wrote our own concurrence homology software in $R$ \linebreak
\cite{RDCT03:RManual}. 
Other software for computing homology include Dionysus  \linebreak
(http://mrzv.org/software/dionysus), Perseus  \linebreak
(www.math.rutgers.edu/$\sim$vidit/perseus.html), and CHomP \linebreak
(http://chomp.rutgers.edu/). 

The distribution of the time needed to compute the homology for each subject had a very long right hand tail. Usually a few hours sufficed to compute the homology, but sometimes a week did not.

\section{``Combinatorial explosion''}  
		\label{S:combo.explosion}
Later (section \ref{S:some.findings}) we look at seventh-order dependence among the regions of the ``default mode network (DMN)'' 
\cite{lqUamcKbbBfxCmpM09.DefaultModeNetwork}, 
in each subject in our fMRI dataset. In our interpretation the DMN consists of 40 regions. 
For each subject we discarded eight regions (section \ref{S:dichotomization}). Thus, we examine seventh-order dependence in a 32-way table.

An agnostic seventh-order log linear analysis would result in $\binom{32}{7} = 3,365,856$ distinct $\lambda$'s for each subject (compared to the 6,144 fMRI BOLD values -- 192 time points in 32 regions -- in each subject's data). The rapid growth in $\binom{V}{p}$ as $p$ increases deserves to be called a ``combinatorial explosion''. (See 
\cite[p.\ 150]{aA90.CatDatAnal}.)

By contrast we found that the data summaries produced by concurrence homology included at most hundreds of numbers per subject, even if ``localization'' (section \ref{S:localization}) was employed. Moreover, those numbers are structured in a way that aids interpretation. Thus, concurrence homology provides parsimonious descriptions of high-order dependence (the cost is in computation time).

\section{Persistence} \label{S:persistence}
In the filtered Curto-Itskov complex for data set I, shown in the top row of figure \ref{F:toy.filtered.conc.cmplxes}, two triangles appear. But the two triangles are related: Moving in decreasing order of frequency level, i.e., from right to left, a triangular hole (labeled ``$\alpha$'') appears (is ``born'') at frequency level 2 and ``persists'' at frequency level 1. We say that the persistent triangle in data set I ``dies in frequency level 0''. In data set II again a triangular hole appears (is ``born'') at frequency level 2, but it ``dies'' in frequency level 1. By a ``persistent (homology) class'' we mean a collection of homology classes (holes) in various frequency levels that are related to each other in this way. 

Hence, instead of saying that the filtered Curto-Itskov complex of data set I has three 1-dimensional holes, we say that it has two 1-dimensional persistent homology classes with lifespans 2 and 1. Identifying persistent classes and their the births and deaths is ``persistent homology'' 
\cite{hEjlH10.CompTopol}.

Recently there has been much interest in using persistent homology for data analysis, e.g., 
\cite{rG08.PersistentTopologyOfData} and 
\cite{gC09.TopolData}. 
In particular, persistent homology has been applied to brain data \linebreak
\cite{hLmkChKbnKdsL11.DZeroPersHomolFromPET,mkCpBptK09:CorticalPersistence}.
However, concurrence topology appears to be a new method.

Table \ref{Ta:persistence.of.toy.data} shows that births and deaths of all persistent classes in data sets I, II, and III. Note that the persistent homology of the Curto-Itskov complexes discriminates the three datasets. (concurrence homology in dimension 0 registers second-order dependence. It tracks, not holes, but clusters, or, more precisely, connected components, and is akin to single linkage cluster analysis, 
\cite{bsEsLmLdS11.ClusterAnalysis}. The roles of ``$X$'' and ``$Z$'' can be reversed.)

   \begin{table}[h]
	\begin{tabular}{ p{.7in} || p{.35in} | p{.35in} || p{.35in} | p{.35in} || p{.35in} | p{.35in} || p{.75in} |  }
	& \multicolumn{6}{c||}{Data set} \\
	 & \multicolumn{2}{c||}{I} & \multicolumn{2}{c||}{II} & \multicolumn{2}{c||}{III} \\ \cline{2-7}
	 dimension & birth & death & birth & death & birth & death & comp./hole \\  \hline   
	                     & 2 & 0 & 5 & 0 & 5 & 0 & X \\ \cline{2-7} 
	               0      &  &  & 5 & 2 & 5 & 2 & Z \\ \cline{2-7}
	                     &  &  & 4 & 2 & 4 & 2 & Y \\ \hline\hline  
	               1     & 2 & 0 & 2 & 1 & \multicolumn{2}{c||}{(not present)} & $\alpha$ \\  \hline
	               1     & 1 & 0 & 1 & 0 & 1 & 0 & $\beta$  \\ \hline 
	\end{tabular}
\caption{Persistent homology of the filtered complexes in figure \ref{F:toy.filtered.conc.cmplxes} in dimensions 0 and 1. The column ``comp./hole'' identifies the component (dimension 0) and hole (dimension 1) whose births and deaths are listed. \vspace{12pt}}
\label{Ta:persistence.of.toy.data}
  \end{table}

\begin{figure}
      \epsfig{file = 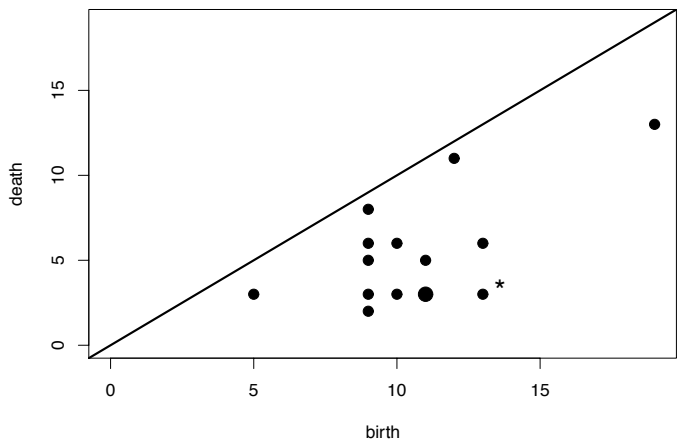, height = 3.1in, } 
      \caption{Dimension 1 persistence plot for the fMRI BOLD values in the time domain in the DMN for subject ``sub01912''. The larger circle indicates two coinciding points.  The dot near the asterisk represents an interesting persistent class discussed in section \ref{SSS:DMN.TD.D1.shrt.cycs}.}    \label{F:one.subj.DMN.TD.D1.pers.plot}
	\end{figure}

Plotting $death$ vs.\ $birth$ yields a ``persistence plot'' for each dimension $d$. (Since we index the filtered Curto-Itskov complex by frequency level, our ``persistence plot'' is different from, but trivially equivalent to, the standard ``persistence diagram'', 
 \cite[p.\ 152]{hEjlH10.CompTopol}.)
Figure \ref{F:one.subj.DMN.TD.D1.pers.plot} shows the persistence plot in dimension 1 (third- and higher-order dependence by equation \eqref{E:dim.d.order.d+2}) for the regions in the DMN
for control subject ``sub01912''. Note that the format of the plot is the same no matter what order of dependence is portrayed. (Plots like this can be averaged over groups.)

Thus, e.g., the dot marked by an ``*'' indicates a persistent \linebreak
1-dimensional homology class that is born in frequency level 13 and dies in frequency level 3. So as one moves downward from frequency level 13 to 3, the frames include increasing numbers of simplices, but this hole is not filled in until frequency level 3. One expects that classes like this one, with a long lifespan, are less likely to appear by chance and are more likely to reflect negative, rather than merely weak, association among the variables. 

It turns out that, indeed, the classes similar to that corresponding to the dot indicated by the ``*'' appear in most subjects in the fMRI data. Investigating this led us to find one of several ways of using concurrence homology to discriminate ADHD subjects from controls (section \ref{SSS:DMN.TD.D1.shrt.cycs}).
   
\section{Concurrence topology in the Fourier domain}
High-order spectral analysis of multivariate time series is a well-studied subject 
\cite{bBejPamZ95.HighOrderSignalProcessing}. There is a concurrence topology version of this.
For each subject the fMRI BOLD data consist of a multivariate time series with one component per region. Concurrence topology of fMRI in the ``time domain'' is carried out by constructing the filtered Curto-Itskov complex from direct dichotomization of BOLD values (section \ref{S:dichotomization}) and treating every time point as a separate observation.
In the ``Fourier domain'', instead of dichotomizing the BOLD signal itself, one dichotomizes the periodograms 
\cite{drB01.BrillingersTimeSeriesBook} of the component series.

Define concurrence in the Fourier domain just as in time domain, but treating angular frequencies as separate observations. This allows the study of high-order dependence while taking into account the time series nature of fMRI. 

\section{fMRI data} \label{S:fMRI.data} 
The fMRI data set was generated at New York University and distributed as part of the 1000 Functional Connectomes project \linebreak
(http://fcon\_1000.projects.nitrc.org/). At the time we began our work this was the largest publicly available resting state fMRI dataset containing clinical data available. This data set includes 41 healthy controls 
(``NewYork\_a\_part1'') and 25 adults diagnosed with ADHD \linebreak
(``NewYork\_a\_ADHD'').

The samples were highly imbalanced with respect to age and gender. Only 20\% of the ADHD group was female, while about half of the controls were. About 25\% of the controls were children (younger than 20; median age = 12), while there were no children in the ADHD group. Among adults, ages ranged from about 21 to about 50 in each group. The median age in the ADHD group was 37, while in the control group the median adult age was 27.

We computed BOLD values for 92 regions, including 40 in the DMN. 
Prior to applying concurrence homology we dropped some regions in a subject-wise fashion (section \ref{S:dichotomization}).

\section{Data analysis of fMRI data}  \label{S:data.analysis}
For each subject we computed summaries of the  persistent homology of the filtered Curto-Itskov complex based on his/her dichotomized fMRI BOLD data or periodograms and compared the distribution of those summaries between groups (or, in one instance, between genders). Thus, we performed inference between subjects, not within subjects. 
Our purpose in this study is to develop methods for using concurrence homology in fMRI data. If a method revealed something of interest in the fMRI data (usually group differences) then we took that as an indication that the method might be a promising one for use elsewhere. 

Thus, the analyses we undertook were exploratory. Operationally, to ``reveal something of interest in the fMRI data'' meant finding an effect that was significant at the $\alpha = 0.05$ level in an appropriate test. (We used Wilcoxon rank sum and chi squared tests and generalized least squares (GLS), 
\cite{jcPdmB00.lme}.) ``Statistical significance'' was merely a flag that indicated analytical methods that might be worthwhile for future use. 

Unless stated otherwise, \emph{all findings we mention concerning the fMRI dataset are statistically significant in this operational, uncorrected \linebreak
sense.} Because our analyses are only exploratory, to save space we omit many details of the analyses performed.

For each subject we computed persistent homology (in both the time and Fourier domains) in dimensions 0 through 5 (corresponding to dependence orders 2 through 7) in the DMN. We also computed persistent homology (in time and Fourier domains) in dimensions 0 through 2 in the whole brain. 
In some cases we also computed the corresponding localization (section \ref{S:localization}) and/or the ``Euler characteristics'' for each subject. (The ``Euler characteristic'' is a one number summary of the homology of a shape 
\cite{{jrM84},dsR08.EulersGem}.)

Since the fMRI dataset is quite imbalanced with respect to age and gender (section \ref{S:fMRI.data}), we sometimes analyzed only the data in adults and/or controlled for age and/or gender.  

In some analyses we summarized the main features of a persistence plot by nine ``moments'':  The first ``moment'' was the number of points in the plot (counting multiplicity). The other ``moments'' are, for $i, j = 0, 1, 2$ (not both 0), 
$\bigl[ \text{average of } (birth^{i}) (lifespan^{j}) \bigr]^{1/(i+j)}$. For the DMN we computed persistent homology in dimensions 0 through 5. Hence, for the DMN we obtained for each subject $6 \times 9 = 54$ moments. For the whole brain we computed persistent homology in dimensions 0 through 2 so each subject has a $3 \times 9 = 27$ moments in the whole brain. We analyzed these multivariate summaries using GLS with $moment$ as the response variable.

\section{Some findings}  \label{S:some.findings} 
Using the GLS analysis just described we picked up  group differences in the DMN in the time domain in dimensions 4 and 5 and in the whole brain in the Fourier domain.

The group difference in the DMN in the time domain in dimensions 4 was a robust finding, in the sense that it manifested itself in a number of analyses. 
The essence of the difference is that fewer ADHD subjects (64.0\%) had any homology in the time domain in the DMN in dimension 4 (i.e., only 64\% had any 4-dimensional holes; this represents $6^{th}$-order dependence by equation \eqref{E:dim.d.order.d+2}) than did controls (92.6\%). 

In the DMN in the Fourier domain the Euler characteristic of the frame in frequency level 1 is typically higher among the ADHD subjects (mean = 1.68, SD = 2.53) than it is among the controls (mean = 0.415, SD = 1.12), another robust finding.

As an informal analysis, we observed in some experiments that the homology one gets from simulated data in which all the regions function independently of each other is far different from what one finds in the real fMRI data. Obviously, brain regions do not function independently of each other. It is reassuring that concurrence homology recognizes this in the data.

We describe further findings concerning the fMRI data set in section \ref{SS:localization.in.fMRI.data}.

\section{Localization} \label{S:localization}
``Localization'' offers a higher resolution description of the topology of the filtered Curto-Itskov complex. Having found a hole (i.e., homology class), it is natural to ask what variables (regions, in our case) are involved? Existence of a hole in the filtered complex requires the cooperation of all variables, but some variables are more directly involved than others. 

For simplicity consider dimension $d = 1$ (third-order dependence, by equation \eqref{E:dim.d.order.d+2}). In dimension 1, a ``cycle'' (``1-cycle'') is a union of one or more closed polygons made up of 1-simplices in the complex. (In general, a $d$-cycle is a union of closed polyhedra made up of $d$-simplices in the complex.) E.g., using an obvious notation, in figure \ref{F:toy.filtered.conc.cmplxes}, data set I, $XY+YZ+ZX$ is a 1-cycle. That cycle is ``short'' because it consists of just three 1-simplices. In general, a ``short'' $d$-cycle is one consisting of $d+2$ simplices, the smallest number that can form a $d$-cycle. 

A homology class consists of cycles that wrap around one or more holes. All holes are surrounded by cycles. But not all holes are surrounded by \emph{short} cycles. E.g., in figure \ref{F:toy.filtered.conc.cmplxes} again, there are no short cycles
that surround the hole marked ``$\beta$''. Conversely, even if a homology class has a short cycle, in general it will also have cycles that are not short. By ``localization'' of a hole, we mean finding \emph{all} short cycles, but only short cycles, that surround that hole, if there are any. 
(
\cite{tkDaHbK08.OptimalHomolCycles} discusses a different notion of localization.) We computed the localization of \emph{all} holes in various dimensions in all subjects. (For this paper, localization was carried out separately for each frequency level, i.e., persistence of homology classes was ignored in the localization.)

\subsection{Localization in the fMRI data}  \label{SS:localization.in.fMRI.data}
\subsubsection{Dimension 1 in the DMN in the time domain}  \label{SSS:DMN.TD.D1.shrt.cycs} 
A short one dimensional cycle involves three regions. In the DMN in the time domain we found  
7,427 distinct short 1-cycles across all subjects. (There are 40 regions in the DMN. $\binom{40}{3} = 9,880$ distinct short cycles are theoretically possible for a single subject; median number of distinct short 1-cycles per subject = 260.) One subject has a one-dimensional homology class (hole) containing 164 short 1-cycles in a single frequency level. 

We select the most important short cycles using two criteria. The first is the number of subjects having the cycle and the second is the lifespan of the cycle. A cycle may represent homology across a range of frequency levels. The ``lifespan'' of the cycle is the number of frequency levels in which it does so. The lifespan of a cycle can never be longer than that of the persistent homology class to which it belongs.

The short cycle whose homology class persistence is plotted at  the point marked by ``*'' in figure \ref{F:one.subj.DMN.TD.D1.pers.plot} appears in 13 subjects and, for subject ``sub01912'', has cycle lifespan = 8. Call this cycle $z$. In subject ``sub01912'' this triplet of regions is well connected at second-order, but, comparatively speaking, not even indirectly well connected at order 3. 

We performed an analysis under the null hypothesis that all possible 9,880 triplets of default mode regions are equally likely to be short cycles in a given subject. We assumed that short cycles were selected from the 9,880 independently \emph{between} subjects, but not necessarily independently \emph{within} subjects. Then, based on a simple model, a heavily Bonferronized upper bound on the probability that \emph{some} triplet will be a short cycle for 13 or more subjects is 0.021. Thus, $z$ is rather special.

Now, presence of $z$ itself does not differentiate the ADHD and control groups, but the 29 short cycles that wrap tightly around the same hole that $z$ does in subject ``sub01912'' \emph{do} distinguish the groups. 

We can refine this. 16 of the 29 short cycles appear at least twice each in each diagnostic group. 
19 out of 25 ADHD subjects (76\%) have at least one of the 16 short cycles, but only 18 out of 41 controls (44\%)  have any. This difference was another of our robust findings.

The frequencies of occurrence of each of the 13 regions involved in any of the 16 short cycles are very similar in the two groups. Neither do the groups differ in frequency of occurrence of any particular short cycle among 16. It appears that there is a particular hole or family of related holes that occur in many of the subjects' filtered Curto-Itskov complexes. We are detecting a subtle more or less reproducible feature in the data that presumably would be common in people in the general control and ADHD populations, but more commonly in the latter.

\subsubsection{Dimension 4 in the DMN in the time domain}  \label{SSS:DMN.TD.D4.shrt.cycs}
In dimension $d=4$, a short cycle involves six regions. Out of $\binom{40}{6} = 91,390$ theoretically possible 4-dimensional short cycles in the DMN-time domain 1,497 appear in the data. The median number of distinct short 4-cycles per subject is 12.5. 

Call a class ``narrow'' if it has at least one short representative cycle. Thus, the holes marked ``$\alpha$'' in figure \ref{F:toy.filtered.conc.cmplxes} are narrow, while the holes marked ``$\beta$'' are not.  Homology classes can be summed 
\cite[Chapter 1]{jrM84}. Say that two narrow classes are ``adjacent'' if their sum is also narrow. (The holes corresponding to the adjacent classes do not actually have to be next to each other in space.)

The presence of adjacent pairs of classes in dimension 4 does not discriminate the diagnostic groups, but it does discriminate genders: Only 1 out of the 25 females have any adjacent class pairs, but 13 out of the 41 males do.

\subsubsection{Dimension 2 in the whole brain in the Fourier domain} \label{SSS:dim2.WB.F.domn}
There are 92 regions in the ``whole brain''. Out of $\binom{92}{4} = 2,794,155$ distinct theoretically possible 2-dimensional short cycles in the whole brain and Fourier domain, 7,933 appear in the data. The median number of distinct short 2-cycles per subject = 57.5. (A short 2-cycle involves four regions.)

The ``corpus callosum'' consists of white matter and until recently only grey matter was believed to produce a BOLD signal \linebreak
\cite{elMsdBjrGkdBcvBrcD'A10.CorpusCallosumfMRI}. 
However, the corpus callosum region that appears in the fewest number of 2-cycles appears 909 times, which is more often than any non-corpus callosum region appears and much larger than the median number of times (249) that non-corpus callosum regions appear. Of the 2,794,155 possible quadruplets of whole brain regions, 20\% include a region from the corpus callosum, but of the distinct short 2-cycles in the data 65\% include a callosum region. 

Thus, the corpus callosum frequently takes part in quadruplets that are weakly connected at fourth-order. However, \emph{positive} dependence at third-order is needed in order to form a short 2-cycle.  So the five corpus callosum regions cannot be said to be weakly functionally connected to other regions in general. (See remark \ref{R:high.dim.holes}.)

\section{Dichotomization} \label{S:dichotomization} 
Concurrence topology is designed for binary data. The fMRI BOLD signal is continuous.  
For each region we determined at which time points the region is ``active'' and at which it is ``inactive'' by dichotomizing fMRI BOLD values. There is no single level of fMRI BOLD that demarcates activity from inactivity, because fMRI BOLD levels in different regions are incomparable. So a separate threshold is needed for each region (in each subject).

A potential complication is that in some cases dichotomizing can merely amplify noise. Brain functional connectivity means covariation. Without variation there is no covariation. 
The little variation shown by a nearly constant activity level is liable to be noise. Dichotomizing such a slightly varying noise series will amplify it and introduce a noisy binary component in the multivariate series. 

Therefore, in the fMRI data, for \emph{each subject separately} we discarded the 20\% least variable regions. So different subjects may have different regions dropped. (One subject had fMRI BOLD values of 0 for all time points in two regions, likely due to either missing or inaccurate automated labeling of the regions.  For that subject those two regions were also dropped.) This was done separately for the whole brain and DMN. We measured variability by a robust version of the coefficient of variation: 
interquartile range divided by median.  

Whether or not our reasoning in favor of dropping the least variable regions is sound, it is expedient: If all regions are included, the computation of homology takes much longer than if low variability regions are dropped. 

We stress the analysis does not start \emph{after} the 20\% least variable regions are dropped. Dropping the least variable regions is the \emph{first step} in the analysis. So this step does not compromise the agnostic nature (section \ref{S:intro}) of our method. The distribution of regions that were dropped did not differ between the ADHD and control groups, but did depend on age and sex.

In the time domain, separately for each subject and region retained for that subject, we deemed the 20\% (39) time points at which the fMRI BOLD value was highest as ``active''. In the Fourier domain, for each subject and region we set the threshold at the $90^{th}$ percentile of power, because the fMRI BOLD time series had low power in about the highest half of the Fourier frequencies and 20\% of half the Fourier frequencies is the same as 10\% of all of them.

\section{Discussion and Conclusions}
Concurrence topology is a nonparametric method for describing high-order dependence in dichotomous data. We described a particular approach to concurrence topology, called ``concurrence homology''. Using an fMRI data set as a test bed, we explored a number of different ways of deploying concurrence topology. These included persistence, Euler characteristics, and several different ways of mining localizations and found numerous interesting apparent structures in the data.  These findings are only exploratory but we intend to try to replicate our findings in an appropriate independent data set. 

The thresholds used in dichotomization (section \ref{S:dichotomization}) are tuning constants of the method. Our choices are based on informal experiments on a smaller data set independent of our fMRI data set. More experimentation with tuning constants is needed, but is difficult because of the lengthy computing times.

Concurrence homology is computationally intensive but, with that proviso, concurrence topology can be applied, not just to fMRI BOLD data, but to any multivariate dichotomous data. Moreover, we are confident that improved software will greatly expand the range of data that can be analyzed using concurrence homology.

\end{document}